\documentclass[aps,prl,showpacs,twocolumn]{revtex4}
\usepackage{hyperref}

\begin{document}

\title{Maximum elastic deformations of compact stars with exotic equations
of state}
\author{Benjamin J. Owen}
\affiliation{
Center for Gravitational Wave Physics,
Institute for Gravitational Physics and Geometry, and
Department of Physics,
The Pennsylvania State University, University Park, PA 16802-6300}
\date{Received 19 November 2004; revised 13 April 2005}

\preprint{IGPG-04/11-8}

\begin{abstract}

I make the first estimates of maximum elastic quadrupole deformations
sustainable by alternatives to conventional neutron stars.
Solid strange quark stars might sustain maximum ellipticities
(dimensionless quadrupoles) up to a few times $10^{-4}$ rather than a few
times $10^{-7}$ for conventional neutron stars, and hybrid quark-baryon or
meson-condensate stars might sustain up to $10^{-5}$.
Most of the difference is due to the shear modulus, which can be up to
$10^{33}$~erg/cm$^3$ rather than $10^{30}$~erg/cm$^3$ in the inner crust of
a conventional neutron star.
Maximum solid strange star ellipticities are comparable to upper limits
obtained for several known pulsars in a recent gravitational-wave search by
LIGO.
Maximum ellipticities of the more robust hybrid model will be detectable by
LIGO at initial design sensitivity.
A large shear modulus also strengthens the case for starquakes as an
explanation for frequent pulsar glitches.

\end{abstract}


\maketitle

The LIGO Science Collaboration recently used data from the second LIGO
science run (S2) to set upper limits on gravitational wave emission from 28
known pulsars, 9 of which have no competing upper limit from radio
observations~\cite{Abbott:2004ig}.
For those 9 pulsars the best S2 upper limits on neutron star ellipticity [a
dimensionless quadrupole moment, see Eq.~(\ref{epsdef})] are a few times
$10^{-5}$.
With better data now being analyzed, LIGO will soon be sensitive to
ellipticities of $10^{-6}$ or less.
This raises the question of when a detection might be possible, or when
enough non-detections (upper limits) begin to confront some theoretical
models of dense matter.
The answer depends on the maximum ellipticities sustainable in those
models.
I make the first estimates of maximum ellipticities for several exotic
matter models and find that the LIGO S2 search was already sensitive to the
upper end of the theoretical range.
LIGO observational results are becoming astrophysically interesting years
sooner than previously expected.

The maximum elastic deformation sustainable by a neutron star has been
addressed several times in the past few
decades---see~\cite{Ushomirsky:2000ax}, and references therein.
A conventional neutron star consists of a liquid nuclear-matter core
covered by a thin solid crust, which is responsible for the deformation and
whose microphysics can be extrapolated conservatively from laboratory
nuclear physics.
More exotic models of compact stars have been proposed, some including
large solid cores (see~\cite{Glendenning:book} for a summary), but the
maximum deformation has not been quantitatively addressed.
Historically the problem was of interest first in relation to the
``glitch'' phenomenon in pulsars, which was believed to be related to
starquakes~\cite{Ruderman:1969}.
However, the total elastic energy stored in a maximally strained crust is
far too low to explain the strength and frequency of the glitches of the
pulsar Vela~X-1~\cite{Baym:1971}.
Occasionally works on exotic compact stars have mentioned that solid cores
might revive the starquake glitch mechanism, but without estimating
numbers.

In this Letter, I estimate maximum elastic deformations sustainable by
exotic alternatives to neutron stars and work out the implications for
gravitational-wave emission and pulsar glitches.
Of the models extant in the literature, solid strange stars allow the
largest ellipticities---up to $10^3$ times those of neutron
stars---although this model is highly speculative.
Hybrid quark-baryon stars and stars with charged meson condensates, both
based on more robust theories, might allow ellipticities up to a few times
$10^1$ more than those of conventional neutron stars.
This makes detectable gravitational wave emission a prospect for initial
LIGO rather than advanced LIGO and makes the starquake model of glitches
viable again.

There are several sources of uncertainty in such estimates.
The largest is the matter model itself---maximum ellipticities vary by
$10^3$ between conventional neutron stars and solid strange stars.
The second largest is the breaking strain.
I quote fiducial numbers for a breaking strain of $10^{-2}$, which is near
the maximum for terrestrial alloys and may be favored by observations of
neutron stars in low-mass x-ray binaries, but the breaking strain could be
lower by $10^2$--$10^3$~\cite{Ushomirsky:2000ax}.
For hybrid and condensate stars, charge screening might bring the maximum
ellipticity down to that of a neutron star.
These uncertainties justify making several approximations which simplify
the calculations at a cost of introducing relatively small errors as
in~\cite{Ushomirsky:2000ax}.
Relativistic gravity and rotational effects can change the density profile
of a star by tens of percent; but they cancel to some extent and are
smaller than the effect of varying the star's mass a few
percent~\cite{Glendenning:book}, and thus I neglect them.
Because of the high Fermi energies involved, finite temperature plays a
negligible role in determining the maximum ellipticity.
In the strange and hybrid stars, a normal solid crust is believed to be
still present, but its contribution to the ellipticity is a few percent
correction to that of the core.
I quote maximum ellipticities including the maximum 200\% contribution from
the self-gravity of the deformation~\cite{Ushomirsky:2000ax, Cutler}, but
that could go down by a factor of 2.
Further calculations, details, and uncertainties will be presented
elsewhere~\cite{Owen}.

{\it Neutron stars.}---Reference~\cite{Ushomirsky:2000ax} computes in its
Eq.~(69) a maximum $m=2$ quadrupole moment for a neutron star using a
chemically detailed model of the crust.
Correcting the definition of shear modulus~\cite{Cutler}, it reads
\begin{eqnarray}
\label{q22-ucb}
Q_{22,\max} &=& 2.4\times10^{38}~\mathrm{g~cm}^2 \left( \sigma_{\max} \over
10^{-2} \right) \left( R \over 10~\mathrm{km} \right)^{6.26}
\nonumber
\\
&& \times \left( 1.4~M_\odot \over M \right)^{1.2},
\end{eqnarray}
where $\sigma_{\max}$ is the breaking strain of the crust.

The quadrupole~(\ref{q22-ucb}) can be converted to the ellipticity $\epsilon
= (I_{xx}-I_{yy})/I_{zz}$ used in gravitational wave
papers~\cite{Abbott:2004ig}:
\begin{equation}
\label{epsdef}
\epsilon = \sqrt{8\pi/15} \, Q_{22} / I_{zz},
\end{equation}
where the $z$-axis is the rotation axis and $I_{ab}$ is the moment of
inertia tensor.
For conventional neutron stars, Bejger and Haensel~\cite{Bejger:2002ty} find
that the approximation
\begin{eqnarray}
\label{I-normal}
I_{zz} &=& 9.2\times10^{44}~\mathrm{g~cm}^2 \left( M \over 1.4~M_\odot
\right) \left( R \over 10~\mathrm{km} \right)^2
\nonumber
\\
&& \times \left[ 1 + 0.7 \left( M \over 1.4~M_\odot \right) \left(
10~\mathrm{km} \over R \right) \right]
\end{eqnarray}
is accurate to a few percent for a variety of equations of state.
Thus we can write the maximum ellipticity of a conventional neutron star as
\begin{eqnarray}
\label{eps-ucb}
\epsilon_{\max} &=& 3.4 \times 10^{-7} \left( \sigma_{\max} \over 10^{-2}
\right) \left( 1.4~M_\odot \over M \right)^{2.2} \left( R \over
10~\mathrm{km} \right)^{4.26}
\nonumber
\\
&& \times \left[ 1 + 0.7 \left( M \over 1.4~M_\odot \right) \left(
10~\mathrm{km} \over R \right) \right]^{-1}.
\end{eqnarray}
For the fiducial values of mass, radius, and breaking strain,
$\epsilon_{\max}$ is $2\times10^{-7}$ ($6\times10^{-7}$ with self-gravity).

The generalization of Eq.~(\ref{q22-ucb}) to arbitrary equations of state
can be obtained by combining Eqs.~(67) and~(64) of
Ref.~\cite{Ushomirsky:2000ax} as
\begin{equation}
\label{q22}
{Q_{22,\max} \over \sigma_{\max}} = \sqrt{32\pi\over15} \int dr {\mu r^3
\over g} \left( 48 - 14U + U^2 - {dU \over d\ln{r}} \right),
\end{equation}
where $\mu$ is the shear modulus (nonzero only in the solid part of the
star), $g$ is the local gravitational acceleration, and
$U=2+d\ln{g}/d\ln{r}$.
The two bounding cases are incompressible matter and infinitely compressible
matter (a point mass).
Note that the latter is equivalent to a conventional neutron star, where the
mass of the crust is a small fraction of the mass of the star.
For a light crust, $U\ll1$ and $g\approx GM/r^2$; for the incompressible
case, $U=3$ and $g=GMr/R^3$.
If $\mu$ is almost constant [or is replaced by an appropriately averaged
value as below Eq.~(68) of Ref.~\cite{Ushomirsky:2000ax}], Eq.~(\ref{q22})
simplifies for an incompressible completely solid star to
\begin{equation}
\label{q22-simple}
Q_{22,\max} = \gamma \mu R^6 \sigma_{\max} /(GM)
\end{equation}
where $\gamma \approx 13$.
Evaluating Eq.~(\ref{q22}) for a conventional neutron star with a thin crust
and liquid core, $\gamma$ becomes about $120\Delta R/R$, where $\Delta
R\approx R/10$ is the thickness of the crust, and thus $\gamma$ is
numerically almost identical.
The appropriately averaged shear modulus from Ref.~\cite{Ushomirsky:2000ax}
is $\mu \approx 4\times10^{29}$~erg/cm$^3$, a factor of a few below its
maximum value at the bottom of the crust.

{\it Solid strange stars.}---The idea that some ``neutron stars'' are in
fact made of strange quarks was proposed in the 1970s~\cite{Bodmer:1971we}.
The idea that such stars are solid currently is being pursued by Xu's
group, beginning with Ref.~\cite{Xu:2003xe}.
(This is distinct from a crystalline color superconducting quark
phase~\cite{Alford:2000ze}, which I do not consider here.)
Xu notes that the burst oscillation frequencies observed in low-mass x-ray
binaries correspond to the first few torsional modes of a solid strange
star---if the matter has a typical shear modulus $\mu \approx
4\times10^{32}$~erg/cm$^3$, a thousand times the typical value in the crust
of a conventional neutron star.
Xu estimates that quarks clustered in groups of 18 or so could produce such
a shear modulus.
Since Ref.~\cite{Xu:2003xe} was published, the burst oscillation frequency
has been observed to closely match the spin frequency of the neutron star in
at least one system~\cite{Chakrabarty:2003kt}.
This renders the identification with torsional mode frequencies
problematic.
However, the x-ray burst oscillation mechanism may be different for
different binaries, and it is worth considering the effect on the maximum
elastic deformation if the shear modulus is very high for whatever reason.

Using Xu's shear modulus in Eq.~(\ref{q22-simple}) gives
\begin{eqnarray}
\label{q22-xu}
Q_{22,\max} &=& 2.8\times10^{41}~\mathrm{g~cm}^2 \left( \mu \over
4\times10^{32}~\mathrm{erg/cm}^3 \right)
\nonumber
\\
&& \times \left( \sigma_{\max} \over 10^{-2} \right) \left( R \over
10~\mathrm{km} \right)^6 \left( 1.4~M_\odot \over M \right).
\end{eqnarray}
Bejger and Haensel~\cite{Bejger:2002ty} find a different empirical formula
for the moment of inertia for strange stars,
\begin{eqnarray}
I_{zz} &=& 1.7\times10^{45}~\mathrm{g~cm}^2 \left( M \over 1.4~M_\odot
\right) \left( R \over 10~\mathrm{km} \right)^2
\nonumber
\\
&& \times \left[ 1 + 0.14 \left( M \over 1.4~M_\odot \right) \left(
10~\mathrm{km} \over R \right) \right].
\end{eqnarray}
This combined with Eq.~(\ref{q22-xu}) yields a maximum ellipticity
\begin{eqnarray}
\epsilon_{\max} &=& 2 \times 10^{-4} \left( \sigma_{\max} \over 10^{-2}
\right) \left( 1.4~M_\odot \over M \right)^3 \left( R \over 10~\mathrm{km}
\right)^3
\nonumber
\\
&& \times \left[ 1 + 0.14 \left( M \over 1.4~M_\odot \right) \left(
10~\mathrm{km} \over R \right) \right]^{-1}
\end{eqnarray}
for solid strange stars, where I have inserted the scalings of $\mu$ from
Ref.~\cite{Xu:2003xe} except for the $f$ and the $x$ dependence, which
roughly cancel out.
With self-gravity, the canonical number is $\epsilon_{\max} =
6\times10^{-4}$.

{\it Hybrid and meson condensate
stars.}---Glendenning \cite{Glendenning:1992vb} showed that the phase
transition from baryonic matter to quark matter occurs over a range of
pressures rather than at a single value.
(The argument holds for stars with charged meson condensates as well as for
stars with quark-baryon cores~\cite{Glendenning:book}.
The numbers are very similar, so I discuss only hybrid stars.)
Purely baryonic matter at high densities is isospin asymmetric, which is
energetically unfavorable.
Moving toward isospin symmetry (creating more protons) would require
negative charges to compensate, and leptons are not favored since they are
nearly massless.
When the quark phase becomes available, baryonic matter can attain positive
charge density by moving negative charge into areas of quark matter.

The crystal structure of the mixed phase changes with density.
Immediately above the threshold density for the beginning of the phase
transition, the mixed phase consists of small quark droplets arranged in a
bcc lattice in a baryonic background.
As the density increases, the droplets grow and merge to become rods, then
slabs.
Eventually the baryonic matter becomes the minority slabs, then rods, then
finally droplets before disappearing entirely.
The locations of these layers are highly parameter-dependent; as an upper
limit the mixed-phase crystal can occupy the innermost 8~km of the
star~\cite{Glendenning:book}.

The shear modulus of a bcc lattice of point charges can be written in the
parameters of Ref.~\cite{Glendenning:book}
as~\cite{Baym:1971,Ushomirsky:2000ax}
\begin{equation}
\mu = 0.075\, q^2 D^6 / S^4.
\end{equation}
I have assumed spherical droplets of (esu) charge density $q$, diameter $D$,
and spacing $S$ (Wigner-Seitz cell diameter, or $\sqrt{3}/2$ times the
lattice constant).
Corrections due to the nonsphericity of the droplets reduce this by an
amount that is small in most of the layer.
Typical numbers from Chap.~9 of Ref.~\cite{Glendenning:book} give
\begin{eqnarray}
\label{mu-big}
\mu &=& 4\times10^{32}~\mathrm{erg/cm}^3
\nonumber
\\
&& \times \left( q \over -0.4e/\mathrm{fm}^3 \right)^2 \left( D \over
15~\mathrm{fm} \right)^6 \left( 30~\mathrm{fm} \over S \right)^4.
\end{eqnarray}
This is of order $10^3$ times the typical value in the inner crust,
mainly due to the charge of the droplets (about $10^3$ rather than $Z<55$ in
the crust) although the density of droplets is slightly greater
too~\cite{Douchin:2001sv}.

The dominant correction to Eq.~(\ref{mu-big}) is due to charge screening.
This effect is difficult to evaluate precisely, but a rough estimate can be
made as follows.
Heiselberg, Pethick, and Staubo~\cite{Heiselberg:1992dx} estimate the
screening length
\begin{equation}
\label{lambda}
\lambda = \left[ 4\pi \sum_i Q_i^2 \left( \partial n_i \over \partial \mu_i
\right) \right]^{-1/2}
\end{equation}
in the mixed phase as 10~fm in the baryonic matter and 5~fm in the quark
matter.
(Here $Q_i$ is the charge of species $i$, $n_i$ is its number density, and
$\mu_i$ is its chemical potential.)
Detailed calculations~\cite{Lindblom:2001hd} of the partial derivatives in
Eq.~(\ref{lambda}) for baryonic matter without a quark phase suggest that
$\lambda \approx 5$~fm is a lower limit.
Since these lengths are comparable to the droplet size and separation,
screening will appreciably reduce electrostatic effects but not make them
negligible.
(The leptons can be neglected since their charge density is
tiny~\cite{Glendenning:book}.)

First note that screening does not appreciably change the droplet size.
The quark volume fraction $\chi\approx(D/S)^3$ is set by, e.g., the
pressure, and $D$ is found by minimizing the sum of surface and Coulomb
energy densities at fixed $\chi$~\cite{Glendenning:book}.
The mean charge density $\bar{q}$ is then fixed even under rearrangement of
charges due to screening.
The Coulomb energy density can be written $g(D/\lambda)C(\chi)D^2$, where
$g$ is a geometric factor.
Going from uniform density ($D/\lambda\ll1$) to a shell of charge
($D/\lambda\to\infty$) only reduces $g$ by 1/6.
Using a rough approximation $g\propto5 + \exp(-D/\lambda)$, the screened $D$
is related to the unscreened $D_0$ by
\begin{equation}
D_0^3 = D^3 + D^4/(10\lambda) \exp(-D/\lambda).
\end{equation}
Since $D_0$ and $\lambda$ are comparable, this is a few percent correction
at most, and since $(D-D_0) \ll \lambda$, even the exponential factor is
corrected only by a few percent.

Given that the droplet size does not change appreciably, and given that the
charges inside the droplet rearrange themselves respecting spherical
symmetry, the problem outside the droplet reduces to the classic screening
problem.
The potential is multiplied by the Yukawa factor $\exp(-r/\lambda)$.
Since the shear modulus is roughly a second derivative of the potential
energy, it is multiplied by roughly this factor.
For typical $D$ and $S$ values, screening then reduces the shear modulus by
$e^3\approx20$ ($\lambda \approx 10$~fm) or $e^6\approx400$ for $\lambda
\approx 5$~fm.
The upper limit (weak screening) on the shear modulus is then $\mu \approx
2\times10^{31}$~erg/cm$^3$.

The effective shear modulus for the rod and slab configurations can be
estimated from the droplet result.
Matter made of rods cannot resist a shear stress along the axis of the rods,
but will have a perpendicular response similar to that of the droplets.
This anisotropic case requires an elastic modulus tensor rather than a shear
modulus scalar.
However, if the glitch history of the neutron star has led to
granulation~\cite{Glendenning:book}, the formation of small domains with
different principal directions, then a macroscopic rms response of the
matter averaged over many domains is isotropic with an effective shear
modulus reduced by $\sqrt{2/3}$ or $\sqrt{1/3}$, which can be neglected
here.

Now evaluate Eq.~(\ref{q22}).
The density of the core of the star varies only by a factor of a few, so use
the incompressible limit.
Most of the integral comes from the droplet and rod layers, where the
weak-screening shear modulus is roughly constant at
$2\times10^{31}$~erg/cm$^3$.
Then
\begin{equation}
{Q_{22,\max} \over \sigma_{\max} /10^{-2}} = 3.5\times10^{39}\mbox{
erg/cm}^3 \left( 1.4~M_\odot \over M \right) \left( R_c \over 8\mbox{ km}
\right)^6.
\end{equation}
where $R_c$ is the radius of the hybrid core.
Bejger and Haensel~\cite{Bejger:2002ty} find that hybrid stars obey the same
moment of inertia relation~(\ref{I-normal}) as normal neutron stars, so
\begin{eqnarray}
{\epsilon_{\max} \over 5\times10^{-6}} &=& \left( \sigma_{\max} \over
10^{-2} \right) \left( 1.4~M_\odot \over M \right)^2 \left( R_c \over
8\mbox{ km} \right)^6 \left( 10\mbox{ km} \over R \right)^2
\nonumber
\\
&& / \left[ 1 + 0.7 \left( M
\over 1.4~M_\odot \right) \left( R \over 10\mbox{ km} \right) \right],
\end{eqnarray}
for a fiducial value of $3\times10^{-6}$, or up to $9\times10^{-6}$ with
the self-gravity of the deformation.

{\it Implications.}---What are the immediate consequences for LIGO?
The S2 paper~\cite{Abbott:2004ig} quotes direct gravitational-wave
observational upper limits on $\epsilon$ for 28 pulsars.
However, 19 of these pulsars already have lower indirect upper limits on
$\epsilon$ (typically $10^{-8}$ or less) due to the measured spin-downs.
The remaining 9 pulsars are in globular clusters where the spin-down is
obscured by acceleration, and thus have no competing upper limit.
The S2 upper limits on $\epsilon$ for these are 4--24$\times10^{-5}$, all
within the maximum I estimate for a solid strange star.
With LIGO's upcoming data run at full initial sensitivity, the same 9
pulsars will be observable at $\epsilon$ of 1--8$\times10^{-6}$, within the
maximum for hybrid stars; and the Crab pulsar will be observable at
$\epsilon=1.2\times10^{-4}$, 6 times less than its spin-down limit and
within the solid strange star range~\cite{Cutler:2002me}.
An all-sky search for unknown neutron stars could detect hybrid stars
within a~kpc and solid strange stars at the galactic core with tens of
teraflops computing power~\cite{Cutler:2002me}.

If a pulsar is observed in gravitational waves with $\epsilon \gg 10^{-7}$,
it cannot be a conventional neutron star.
An upper limit (nondetection) at higher $\epsilon$ does not rule out any
exotic model---a given star may happen to be nowhere near its breaking
strain.
However, with enough strict upper limits, population statistics and
deformation mechanisms can be constrained.

Bildsten~\cite{Bildsten:1998ey} proposed that the spin frequencies of stars
in low-mass x-ray binaries are set by equilibrium between accretion torque
and gravitational radiation from thermally induced deformations of the
crust.
In order to match the observed spin frequencies, this requires quadrupoles
$\approx 10^{38}$~g~cm$^2$.
For the exotic stars considered here, such quadrupoles under anisotropic
accretion are possible if the breaking strain is smaller than $10^{-3}$ or
if the detailed accretion physics (temperature dependence, spreading of
material, etc.) prevents achieving breaking strain.

Gravitational waves from freely precessing neutron stars have been
considered poor prospects even for advanced LIGO.
But if internal damping is weak, a population of stars precessing after
birth with $\epsilon=10^{-4}$ would be detectable with broadband advanced
LIGO~\cite{Jones:2001yg}.

A starquake that causes a glitch will also cause a burst of gravitational
waves as the modes of the star are excited and ring down.
This amplitude is determined by the energy in the glitch, which is
determined by the observed frequency jump and thus is not affected by exotic
matter models.
But the maximum elastic energy in the star scales as the shear modulus, and
thus is up to $10^3$ times larger for quark stars than for conventional
neutron stars.
Vela glitches are still too large and frequent (by several orders of
magnitude) to be explained as quakes, but the mean time predicted between
quakes is reduced for the Crab pulsar to a few
years~\cite{Baym:1971}---comparable to what is observed.

After this Letter was submitted, SGR~1806-20 underwent a giant
superflare~\cite{Hurley:2005zs, Palmer:2005mi} with estimated energy more
than $10^{46}$~erg.
Theoretical models equate this energy with the maximum elastic energy of
the star, which is problematic for a normal crust but feasible with the
exotic models considered here.

I thank C.~Cutler, S.~Finn, I.~Jones and M.~Landry for helpful discussions.
This work was supported by NSF grants No.\ PHY-0245649 and No.\ PHY-0114375
(the Center for Gravitational Wave Physics).


\begin{thebibliography}{21}
\expandafter\ifx\csname natexlab\endcsname\relax\def\natexlab#1{#1}\fi
\expandafter\ifx\csname bibnamefont\endcsname\relax
  \def\bibnamefont#1{#1}\fi
\expandafter\ifx\csname bibfnamefont\endcsname\relax
  \def\bibfnamefont#1{#1}\fi
\expandafter\ifx\csname citenamefont\endcsname\relax
  \def\citenamefont#1{#1}\fi
\expandafter\ifx\csname url\endcsname\relax
  \def\url#1{\texttt{#1}}\fi
\expandafter\ifx\csname urlprefix\endcsname\relax\def\urlprefix{URL }\fi
\providecommand{\bibinfo}[2]{#2}
\providecommand{\eprint}[2][]{\url{#2}}

\bibitem[{\citenamefont{Abbott et~al.}(2005)}]{Abbott:2004ig}
\bibinfo{author}{\bibfnamefont{B.}~\bibnamefont{Abbott}}
  \textit{\bibnamefont{et~al.}}
  (\bibinfo{collaboration}{LIGO Scientific Collaboration}),
  \bibinfo{journal}{Phys. Rev.  Lett.} \textbf{\bibinfo{volume}{94}},
  \bibinfo{pages}{181103}
  (\bibinfo{year}{2005}).

\bibitem[{\citenamefont{Ushomirsky et~al.}(2000)\citenamefont{Ushomirsky,
  Cutler, and Bildsten}}]{Ushomirsky:2000ax}
\bibinfo{author}{\bibfnamefont{G.}~\bibnamefont{Ushomirsky}},
  \bibinfo{author}{\bibfnamefont{C.}~\bibnamefont{Cutler}}, \bibnamefont{and}
  \bibinfo{author}{\bibfnamefont{L.}~\bibnamefont{Bildsten}},
  \bibinfo{journal}{Mon. Not. R. Astron. Soc.}
  \textbf{\bibinfo{volume}{319}}, \bibinfo{pages}{902} (\bibinfo{year}{2000}).

\bibitem[{\citenamefont{Glendenning}(2000)}]{Glendenning:book}
\bibinfo{author}{\bibfnamefont{N.~K.} \bibnamefont{Glendenning}},
  \emph{\bibinfo{title}{Compact Stars: Nuclear Physics, Particle Physics, and
  General Relativity}} (\bibinfo{publisher}{Springer}, \bibinfo{address}{New
  York}, \bibinfo{year}{2000}), \bibinfo{edition}{2nd} ed., \bibinfo{note}{and
  references therein}.

\bibitem[{\citenamefont{Ruderman}(1969)}]{Ruderman:1969}
\bibinfo{author}{\bibfnamefont{M.~A.} \bibnamefont{Ruderman}},
  \bibinfo{journal}{Nature (London)} \textbf{\bibinfo{volume}{223}},
  \bibinfo{pages}{597} (\bibinfo{year}{1969}).

\bibitem[{\citenamefont{Baym and Pines}(1971)}]{Baym:1971}
\bibinfo{author}{\bibfnamefont{G.}~\bibnamefont{Baym}} \bibnamefont{and}
  \bibinfo{author}{\bibfnamefont{D.}~\bibnamefont{Pines}},
  \bibinfo{journal}{Ann. Phys. (N.Y.)} \textbf{\bibinfo{volume}{66}},
  \bibinfo{pages}{816} (\bibinfo{year}{1971}).

\bibitem[{\citenamefont{Cutler}(2005)}]{Cutler}
\bibinfo{author}{\bibfnamefont{C.}~\bibnamefont{Cutler}}
  (\bibinfo{howpublished}{private communication}).

\bibitem[{\citenamefont{Owen}(in preparation)}]{Owen}
\bibinfo{author}{\bibfnamefont{B.~J.} \bibnamefont{Owen}}
  (\bibinfo{year}{to be published}).

\bibitem[{\citenamefont{Bejger and Haensel}(2002)}]{Bejger:2002ty}
\bibinfo{author}{\bibfnamefont{M.}~\bibnamefont{Bejger}} \bibnamefont{and}
  \bibinfo{author}{\bibfnamefont{P.}~\bibnamefont{Haensel}},
  \bibinfo{journal}{Astron. Astrophys.} \textbf{\bibinfo{volume}{396}},
  \bibinfo{pages}{917} (\bibinfo{year}{2002}).

\bibitem[{\citenamefont{Bodmer}(1971)}]{Bodmer:1971we}
\bibinfo{author}{\bibfnamefont{A.~R.} \bibnamefont{Bodmer}},
  \bibinfo{journal}{Phys. Rev. D} \textbf{\bibinfo{volume}{4}},
  \bibinfo{pages}{1601} (\bibinfo{year}{1971}).

\bibitem[{\citenamefont{Xu}(2003)}]{Xu:2003xe}
\bibinfo{author}{\bibfnamefont{R.~X.} \bibnamefont{Xu}},
  \bibinfo{journal}{Astrophys. J.} \textbf{\bibinfo{volume}{596}},
  \bibinfo{pages}{L59} (\bibinfo{year}{2003}).

\bibitem[{\citenamefont{Alford et~al.}(2001)\citenamefont{Alford, Bowers, and
  Rajagopal}}]{Alford:2000ze}
\bibinfo{author}{\bibfnamefont{M.~G.} \bibnamefont{Alford}},
  \bibinfo{author}{\bibfnamefont{J.~A.} \bibnamefont{Bowers}},
  \bibnamefont{and}
  \bibinfo{author}{\bibfnamefont{K.}~\bibnamefont{Rajagopal}},
  \bibinfo{journal}{Phys. Rev. D} \textbf{\bibinfo{volume}{63}},
  \bibinfo{pages}{074016} (\bibinfo{year}{2001}).

\bibitem[{\citenamefont{Chakrabarty et~al.}(2003)}]{Chakrabarty:2003kt}
\bibinfo{author}{\bibfnamefont{D.}~\bibnamefont{Chakrabarty}}
  \textit{\bibnamefont{et~al.}}, \bibinfo{journal}{Nature (London)}
  \textbf{\bibinfo{volume}{424}}, \bibinfo{pages}{42} (\bibinfo{year}{2003}).

\bibitem[{\citenamefont{Glendenning}(1992)}]{Glendenning:1992vb}
\bibinfo{author}{\bibfnamefont{N.~K.} \bibnamefont{Glendenning}},
  \bibinfo{journal}{Phys. Rev. D} \textbf{\bibinfo{volume}{46}},
  \bibinfo{pages}{1274} (\bibinfo{year}{1992}).

\bibitem[{\citenamefont{Douchin and Haensel}(2001)}]{Douchin:2001sv}
\bibinfo{author}{\bibfnamefont{F.}~\bibnamefont{Douchin}} \bibnamefont{and}
  \bibinfo{author}{\bibfnamefont{P.}~\bibnamefont{Haensel}},
  \bibinfo{journal}{Astron. Astrophys.} \textbf{\bibinfo{volume}{380}},
  \bibinfo{pages}{151} (\bibinfo{year}{2001}).

\bibitem[{\citenamefont{Heiselberg et~al.}(1993)\citenamefont{Heiselberg,
  Pethick, and Staubo}}]{Heiselberg:1992dx}
\bibinfo{author}{\bibfnamefont{H.}~\bibnamefont{Heiselberg}},
  \bibinfo{author}{\bibfnamefont{C.~J.} \bibnamefont{Pethick}},
  \bibnamefont{and} \bibinfo{author}{\bibfnamefont{E.~F.}
  \bibnamefont{Staubo}}, \bibinfo{journal}{Phys. Rev. Lett.}
  \textbf{\bibinfo{volume}{70}}, \bibinfo{pages}{1355} (\bibinfo{year}{1993}).

\bibitem[{\citenamefont{Lindblom and Owen}(2002)}]{Lindblom:2001hd}
\bibinfo{author}{\bibfnamefont{L.}~\bibnamefont{Lindblom}} \bibnamefont{and}
  \bibinfo{author}{\bibfnamefont{B.~J.} \bibnamefont{Owen}},
  \bibinfo{journal}{Phys. Rev. D} \textbf{\bibinfo{volume}{65}},
  \bibinfo{pages}{063006} (\bibinfo{year}{2002}).

\bibitem[{\citenamefont{Cutler and Thorne}(2002)}]{Cutler:2002me}
\bibinfo{author}{\bibfnamefont{C.}~\bibnamefont{Cutler}} \bibnamefont{and}
  \bibinfo{author}{\bibfnamefont{K.~S.} \bibnamefont{Thorne}}
  (\bibinfo{year}{2002}), \eprint{gr-qc/0204090}.

\bibitem[{\citenamefont{Bildsten}(1998)}]{Bildsten:1998ey}
\bibinfo{author}{\bibfnamefont{L.}~\bibnamefont{Bildsten}},
  \bibinfo{journal}{Astrophys. J.} \textbf{\bibinfo{volume}{501}},
  \bibinfo{pages}{L89} (\bibinfo{year}{1998}).

\bibitem[{\citenamefont{Jones and Andersson}(2002)}]{Jones:2001yg}
\bibinfo{author}{\bibfnamefont{D.~I.} \bibnamefont{Jones}} \bibnamefont{and}
  \bibinfo{author}{\bibfnamefont{N.}~\bibnamefont{Andersson}},
  \bibinfo{journal}{Mon. Not. R. Astron. Soc.}
  \textbf{\bibinfo{volume}{331}}, \bibinfo{pages}{203} (\bibinfo{year}{2002}).

\bibitem[{\citenamefont{Hurley et~al.}(2005)}]{Hurley:2005zs}
\bibinfo{author}{\bibfnamefont{K.}~\bibnamefont{Hurley}}
  \textit{\bibnamefont{et~al.}},
  \bibinfo{journal}{Nature (London)} \textbf{\bibinfo{volume}{434}},
  \bibinfo{pages}{1098} (\bibinfo{year}{2005}).

\bibitem[{\citenamefont{Palmer et~al.}(2005)}]{Palmer:2005mi}
\bibinfo{author}{\bibfnamefont{D.~M.} \bibnamefont{Palmer}}
  \textit{\bibnamefont{et~al.}}, \bibinfo{journal}{Nature (London)}
  \textbf{\bibinfo{volume}{434}}, \bibinfo{pages}{1107} (\bibinfo{year}{2005}).

\end{thebibliography}
\end{document}